# Directional out-coupling of light from a plasmonic nanowire-nanoparticle junction

Danveer Singh[1], Arindam Dasgupta[1], Aswathy V. G.[1], Ravi P. N. Tripathi[1], G.V. Pavan Kumar[1*]

[1]*Photonics & Optical Nanoscopy Laboratory, Indian Institute of Science Education and Research, Pune – 411008, INDIA*
[*]*Corresponding author:pavan@iiserpune.ac.in*

We experimentally show how a single Ag nanoparticle (NP) coupled to an Ag nanowire (NW) can convert propagating surface plasmon polaritons to directional photons. By employing dual-excitation Fourier microscopy with spatially filtered collection-optics, we show single- and dual-directional out-coupling of light from NW-NP junction for plasmons excited through glass-substrate and air-superstrate. Furthermore, we show NW-NP junction can influence the directionality of molecular-fluorescence emission, thus functioning as an optical antenna. The results discussed herein may have implications in realizing directional single-photon sources and quantum plasmon circuitry.

The ability to control and direct light at nanoscale is an important challenge in nano-optics and nanophotonics. To achieve this, plasmons, which are collective oscillations of light and free-electrons at metal–dielectric interface, have been employed for sub-wavelength propagation and localization of light [1]. Plasmons can reduce the mismatch between the cross-section of an excitation optical beam and the absorption cross-section of emitters such as atoms, molecules, and quantum dots, thus mediating and enhancing light-matter interaction [2]. A variety of plasmonic geometries have been innovated and studied in the context of enhancing light-matter interactions. One such geometry is chemically prepared silver nanowires (Ag NWs) [3–6]. These Ag NWs have atomically smooth surfaces that facilitate sub-wavelength propagation of surface plasmon polaritons (SPPs) over a distance of a few microns [7]. They can be used as plasmonic resonators [8], logic gates [9], spontaneous emission amplifiers [10], single photon sources [11–13], and photon-to-plasmon convertors [14], and have opened up new opportunities in quantum plasmonic circuits [15] and nanoscale quantum optics [16]. In the context of NW-based SPP waveguiding, one of the issues to be addressed is to deterministically convert propagating SPPs into directional photons at a desired location on the nanowire. Such directional out-coupling of light from propagating SPPs of NW can be further utilized to influence enhancement and directionality of spontaneous emission from emitter in the close vicinity.

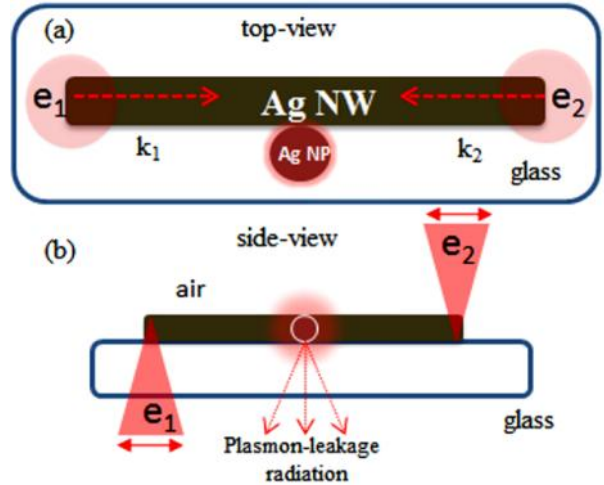

Figure 1: (a) Schematic of the top-view of the experimental configuration of Ag nanowire (NW) coupled to a single Ag nanoparticle (NP), both placed on a glass cover-slip. The regions e1 and e2 are two focused-laser excitations at 633nm that launch SPPs. The double-sided arrows indicate the polarization of excitation.

Motivated by this requirement, herein, we experimentally show how an Ag nanoparticle (NP) in close proximity to Ag NW can out-couple light in directional manner, thereby acting as a nanoscale optical antenna. One of the important observations we make is that the NW-NP junction can out-couple photons in single and dual directions, and further influence directionality of fluorescence emission from molecules in close proximity to NW-NP junction.

First, we discuss the basic principle behind the process of directional light emission from Ag NW-NP junction. Figure 1(a) shows the schematic (top view) of the geometry under study. It consists of an individual Ag NP which is in contact with an Ag NW. Both NW and NP are resting on a glass coverslip.

The SPPs waves on the NW can be optically excited by focusing a laser beam at two distal ends of the NW [Fig. 1(b)]. Such excitation can be achieved by focusing the laser beam either through the glass substrate [excitation $e1$ leading to SPP with wavevector $k1$ in Fig. 1(b)] or through the air medium [excitation $e2$, wavevector $k2$ in Fig. 1(b)]. These excitations can be operated sequentially or simultaneously. The SPPs propagate along the length of the nanowire with a momentum greater than the free space wavelength used for excitation [17]. Therefore, to convert SPPs back to photons at a desired location of NW, momentum has to be subtracted from the surface-bound SPP waves.

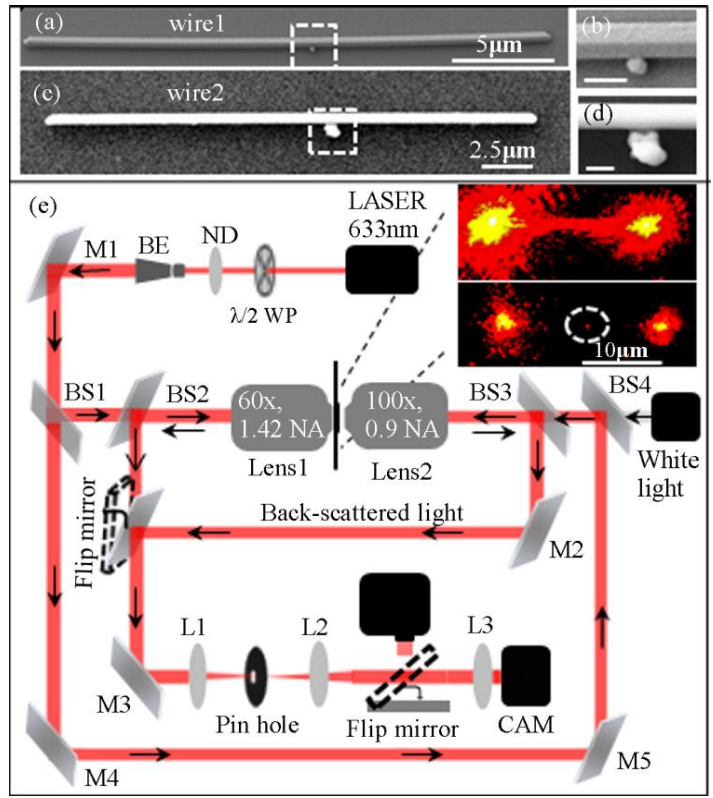

Figure 2: Scanning electron microscopy (SEM) image of NW-NP system resting on glass coverslip at (a) lower magnification with the dashed region showing (b) NW-NP junction (scale bar 200 nm). SEM image of a different NW-NP system imaged at (c) lower and (d) higher magnifications. (e) Schematic of the dual-channel Fourier microscope with spatially filtered collection. M1, M2, M3, M4 and M5 are mirrors; BS1, BS2, BS3 and BS4 are beam splitters; L1, L2 and L3 are lenses; BE-beam expander; ND-neutral density filter; WP -half wave plate; S is sample stage; lens1 is oil immersion objective lens and lens2 is an air objective lens. Upper and lower insets shows an optical image of pristine Ag NW and an Ag NW-NP system excited (633nm) at two distal ends. The dashed circle shows the light emanating from NW-NP junction.

The Ag NP in contact with the Ag NW acts as a defect center to overcome this momentum mismatch, and facilitates the conversion of SPPs to free-space photons at NW-NP junction. One of the aspects of this conversion mechanism is that the scattered light from NW-NP junction is directional in nature, and the directionality of the light can be measured using far-field optical microscopy.

Figures 2(a) and 2(b) shows a representative scanning electron microscopy (SEM) image of NW-NP system resting on glass substrate at two different magnifications. It can be observed that the NW and NP are in contact with each other [Fig. 2(b)]. Such NW-NP systems can be consistently synthesized [see Figs. 2(c) and 2(d) for another NW-NP sample] using polyol method [18], and details of preparation can be found elsewhere [19]. The region of interest for our study is shown as dashed box in Figs. 2(a) and 2(c). This is the NW-NP junction from where the optical signals were spatially filtered and collected using a home-built two-channel optical microscope [Fig. 2(e)]. This microscope had the capability to (i) excite distal ends of the Ag NW either through the glass substrate (lens1) or through the air medium (lens 2) by focusing 633-nm laser light; (ii) spatially filter light from the desired location along the length of the NW [pinhole in Fig. 2(e)]; (iii) capture real- and Fourier-space optical images [L2, L3, CAM in Fig. 2(e)] from spatially filtered region and (iv) record Fluorescence spectra and image from the desired location [spectrometer in Fig. 2(e)]. The inset of Fig. 2(e) shows an optical image of an NW-NP sample excited at two distal ends. The dashed circle in the inset figure shows the out-coupled light from the NW-NP junction. We further studied the directional characteristics of this light with three excitation configurations.

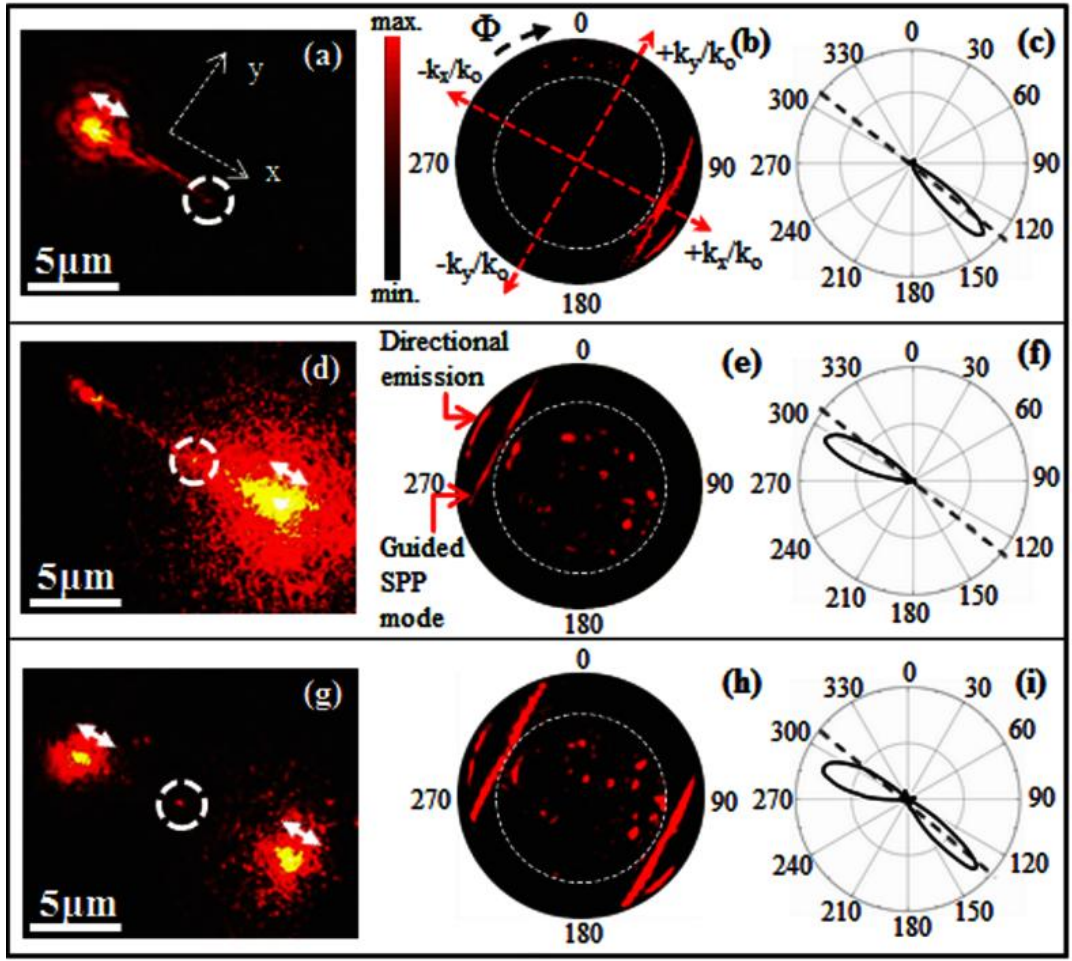

Figure 3:(a) Real-space optical image of Ag NW-NP junction (fig.2a) remotely excited according to e1 excitation in Fig.1. The NW is along x axis. White dashed circle shows the light from NW-NP junction. (b) Fourier-plane ($k_x/k_0$, $k_y/k_0$) intensity (I) image of light from NW-NP junction. The tangential angle is given by $\Phi$ and the radial angle is given by NA of the lens. Inner dashed circle indicates the critical angle at glass-air interface. (c) I Vs $\Phi$ for the arc-emission from NW-NP junction for e1 excitation configuration. (d) Real space image for e2 configuration and corresponding (e) Fourier space optical image from NW-NP junction and (f) I Vs $\Phi$ plot. (g) Real space image for e1+e2 configuration and corresponding (h) Fourier space optical image from NW-NP junction and (i) I Vs $\Phi$ plot. White arrows indicate polarization axis of excitation and dashed lines indicate NW orientation.

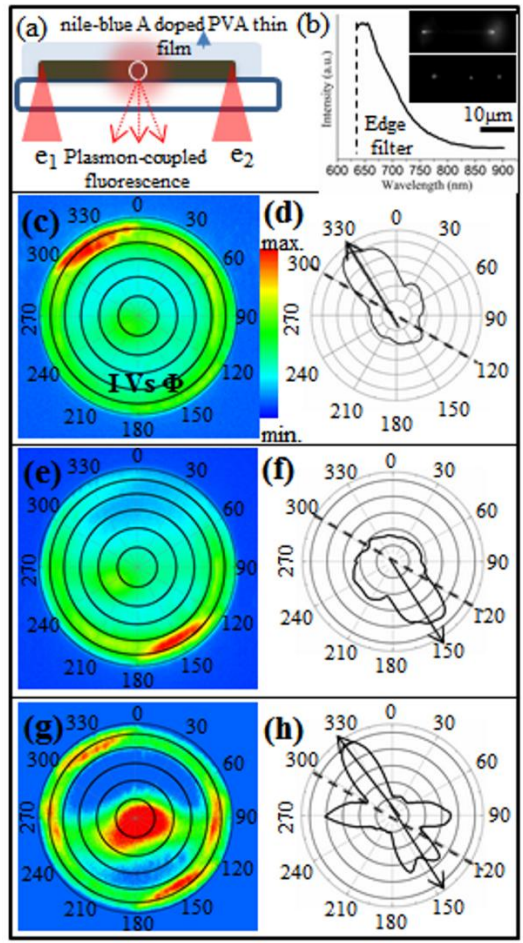

Figure 4: (a) Schematic of the experimental configuration to probe directional fluorescence emission from NW-NP junction. (b) Fluorescence spectra of the emission from the spatially-filtered NW-NP junction, with the upper and lower insets showing fluorescence image of NW (without NP) and NW-NP system, respectively. (c) Fourier-plane optical image from NW-NP junction and (d) I Vs Φ plot for e1 excitation. (e) Fourier space optical image from NW-NP junction and (f) I Vs Φ plot for e2 configuration. (g) Fourier space optical image from NW-NP junction and (h) I Vs Φ plot for counter-propagating e1+e2 excitation. The dashed black-line in polar plots shows the orientation of Ag NW.

*Configuration 1*: Figure 3(a) shows the optical image of an NW-NP system excited through the glass. The polarization of the excitation source was along the length of the NW as shown by the arrow in Fig. 3(a). The dashed circle indicates out-coupled light from NW-NP junction. This spatially filtered region was further imaged in Fourier plane [Fig. 3(b)] of high NA objective lens (1.42NA, 60×). The angular distribution of intensity in the Fourier plane is given by I [Φ, $n \cdot \sin(\theta)$], where tangential angle Φ varies from 0 to $2\pi$, and the radial angle $n \cdot \sin(\theta)$ is the numerical aperture of the lens2 with $n$ presenting the refractive index. The inner circle in the image Fig. 3(b) is of radius 1, indicating the critical angle at glass-air interface, and the outer circle is of radius 1.42 indicating the maximum aperture radius. The NW axis is considered as *x* direction, and *kx*, *ky* represent the wavevectors along *x* and *y* direction, and *k*0 represents the wavevector of the free photons ($k0=2\pi/\lambda$, where $\lambda$ is the free-space wavelength of excitation). Two features were evident in the Fourier-plane image [Fig. 3(b)]. First feature was a straight line for a constant value of *kx/k*0, and the second was an arc-like feature close to the periphery of the outer circle. The straight line indicated the guided mode of the SPP propagating along the length of the NW [17], and the arc indicates the *directional light scattered* from the NW-NP junction. Such directional light scattering has been previously observed only from the distal ends of the NW [20]. Here we have shown, for the first time, directional light emission from a plasmonic NW-NP junction. The arc in Fourier plane was further quantified to determine the spread in the directionality [Fig. 3(c)], and we found the value to be of ~21 degrees. *Configuration 2*: Figure 3(d) shows the optical image of an NW-NP system excited through the air medium. Similar to the previous case, we found that the Fourier plane image showed the guided SPP mode and an arc indicating directional light emission [Fig. 3(e)]. We further quantified the directionality of the emission [Fig. 3(f)] and found a spread of ~24 degrees. *Configuration 3*: Next, we simultaneously excited the distal ends of the Ag NW via glass and air medium [as shown in the optical image of Fig. 3(g)] and captured the light emerging from the NW-NP junction and projected it to the Fourier plane [Fig. 3(h)].

We observed two guided modes and two arcs toward the periphery of the outer circles. The two arcs of light essentially indicated dual channel scattering, and is the central finding of this report. The directionality of the dual channel was further quantified [Fig. 3(i)], and the feature of the emission was found to be a sum of the features observed in Figs. 3(c) and 3(f). All the data in Fig. 3 implied that the NW-NP sub-wavelength junction functioned as a single and dual channel directional scattering point. It is to be noted that parameters such as NW-NP distance, shape of NP, and NW morphology can also influence the directionality. This motivates further studies on such NW-NP systems.

In principle, SPPs in NW couple to NPs leading to directional scattering. Such directional scattering can be further used to excite molecules. Can the NW-NP junction influence the directionality of the fluorescence from emitters? To answer this question, we extrapolated our geometry to the configuration shown in Fig. 4(a). A nano-film (~50 nm thickness) of PVA doped with fluorescent molecules (Nile blue A) was spin-coated over the NW-NP system. Then, we illuminated the ends of the Ag NW through glass substrate and captured the plasmon-coupled fluorescence emission from NW-NP junction. The fluorescence spectra from the NW-NP junction is displayed in Fig. 4(b) with the inset showing the fluorescence image captured from the geometry. The dashed circle in the inset figure shows the fluorescence emission from the NW-NP junction. This fluorescent light was further projected to the Fourier-plane of a high NA lens (as discussed previously) to determine the directionality of the emission. Figure 4(c) shows the Fourier-space image for *e*1 illumination. We found that the fluorescent light collected from the NW-NP junction was indeed directional in nature. We further quantified the directivity of emission and the spread in the angle Φ was found to be ~46 degrees [Fig. 4(d)]. The dashed line in Fig. 4(d) indicates the orientation of the long axis of the NW. Figure 4(e) shows fluorescence emission in Fourier-plane when the other end of NW was illuminated (*e*2 illumination). We found that the directionality in the Fourier-plane was almost opposite [Fig. 4(e)], and the data in Fig. 4(f) showed a spread in the angle Φ to be ~54 degrees. Next we exited the counterpropagating plasmons by exciting both ends of the NW (*e*1+*e*2 illumination), and we recorded the Fourier-space image of the light emerging from NW-NP junction [Fig. 4(g)]. The feature of the emission had two pronounced directions [along 330 and 150 degrees in Fig. 4(h)]. In addition to this, we observed emission along angles 90 and 270 degrees. These emission directionalities with geometrical features of the NW-NP junction need further investigation for complete understanding. All the data in Fig. 4 suggests that NW-NP junction can function as single- and dual-channel optical antenna, which can further influence the directionality of fluorescence emission.

To conclude, we have shown Ag NW-NP junction can be utilized as single- and dual-channel optical antenna that can influence directivity of elastic scattering and fluorescence emission of molecules. As a future prospect, it would be interesting to place single emitters, such as nitrogen-vacancy in diamond [13], close to the NW-NP junction and explore the single-photon emissivity. As the mode volume of the NW-NP junction is expected to be very small, one should expect the Purcell enhancement factor to be very large at such junctions. Such geometries not only enhance the rate of spontaneous emission from emitters, but also influence their directionality.

The authors acknowledge funding from the DST SERB grant (SR/S2/LOP-0007/2011) and the DST Nanoscience Unit Grant (SR/NM/NS-42/2009). GVPK thanks DST for the Ramanujan fellowship, and DS thanks the INSPIRE PhD fellowship.